\pdfoutput=1
\PassOptionsToPackage{table}{xcolor}
\documentclass[sigconf, screen, nonacm]{acmart}

\usepackage{textcomp}
\usepackage{algorithm}
\usepackage{algpseudocode}
\usepackage{array}
\usepackage{tabularx}
\usepackage{tikz}
\usetikzlibrary{arrows.meta}
\definecolor{mqablue}{HTML}{355E7C}
\definecolor{mqateal}{HTML}{00796B}
\graphicspath{{fig/}}

\definecolor{mechanismSurface}{HTML}{BEDBFA}
\definecolor{mechanismHard}{HTML}{FFE09B}
\definecolor{mechanismPreference}{HTML}{A9E2CB}
\definecolor{mechanismRegion}{HTML}{F5BDD4}
\definecolor{mechanismExecution}{HTML}{FFCBA8}
\definecolor{mechanismSelection}{HTML}{D5BDF2}
\definecolor{surfaceInk}{HTML}{1766B2}
\definecolor{hardInk}{HTML}{946000}
\definecolor{preferenceInk}{HTML}{007357}
\definecolor{regionInk}{HTML}{B32661}
\definecolor{executionInk}{HTML}{AD480D}
\definecolor{selectionInk}{HTML}{6D31AE}
\DeclareRobustCommand{\mechanismHighlight}[2]{\begingroup\setlength{\fboxsep}{1pt}\colorbox{#1}{\textcolor{black}{#2}}\endgroup}
\DeclareRobustCommand{\surfaceHighlight}[1]{\mechanismHighlight{mechanismSurface}{#1}}
\DeclareRobustCommand{\hardHighlight}[1]{\mechanismHighlight{mechanismHard}{#1}}
\DeclareRobustCommand{\preferenceHighlight}[1]{\mechanismHighlight{mechanismPreference}{#1}}
\DeclareRobustCommand{\regionHighlight}[1]{\mechanismHighlight{mechanismRegion}{#1}}
\DeclareRobustCommand{\executionHighlight}[1]{\mechanismHighlight{mechanismExecution}{#1}}
\DeclareRobustCommand{\selectionHighlight}[1]{\mechanismHighlight{mechanismSelection}{#1}}
\DeclareRobustCommand{\mechanismTag}[2]{\tikz[baseline=(tag.base)]{\node[circle,fill=#1,text=white,
    inner sep=0pt,minimum size=9pt,font=\fontsize{6.5}{7}\selectfont\bfseries]
    (tag) {#2};}\nobreak\hspace{1.5pt}}
\DeclareRobustCommand{\surfaceTag}{\mechanismTag{surfaceInk}{1}}
\DeclareRobustCommand{\hardTag}{\mechanismTag{hardInk}{2}}
\DeclareRobustCommand{\preferenceTag}{\mechanismTag{preferenceInk}{3}}
\DeclareRobustCommand{\regionTag}{\mechanismTag{regionInk}{4}}
\DeclareRobustCommand{\executionTag}{\mechanismTag{executionInk}{5}}
\DeclareRobustCommand{\selectionTag}{\mechanismTag{selectionInk}{6}}
\definecolor{extDiagInk}{HTML}{00838F}
\definecolor{extProposeInk}{HTML}{6A1B9A}
\definecolor{extImplInk}{HTML}{B35900}
\definecolor{extStaticInk}{HTML}{C62828}
\definecolor{extSearchInk}{HTML}{2E7D32}
\DeclareRobustCommand{\extDiagTag}{\mechanismTag{extDiagInk}{1}}
\DeclareRobustCommand{\extProposeTag}{\mechanismTag{extProposeInk}{2}}
\DeclareRobustCommand{\extImplTag}{\mechanismTag{extImplInk}{3}}
\DeclareRobustCommand{\extStaticTag}{\mechanismTag{extStaticInk}{4}}
\DeclareRobustCommand{\extSearchTag}{\mechanismTag{extSearchInk}{5}}
\DeclareRobustCommand{\ablEvidenceTag}{\mechanismTag{extDiagInk}{A}}
\DeclareRobustCommand{\ablCheckTag}{\mechanismTag{extStaticInk}{B}}
\DeclareRobustCommand{\ablTransferTag}{\mechanismTag{extSearchInk}{C}}
 \definecolor{taxhead}{HTML}{355E7C}
\definecolor{taxopta}{HTML}{E8F0F5}
\definecolor{taxoptb}{HTML}{F5F8FA}
\definecolor{taxrule}{HTML}{8FA7B8}
\newcommand{\tblhead}[1]{\textcolor{white}{\textbf{#1}}}

\newcommand{\systemname}{\textnormal{\textsc{Contour}}}
\newcommand{\papertitle}{Contour: Extensible Mapping Surfaces for Agentic Accelerator Kernel Generation}
\newcommand{\candidatezero}{\textsc{Candidate-0}}
\newcommand{\kernelcraftname}{KernelCraft}
\newcommand{\sysexplore}{Contour-Explore}
\newcommand{\sysext}{Contour-Ext}

\begin{document}

\title{\papertitle}

\newcommand{\affmark}[1]{\texorpdfstring{$^{#1}$}{}}
\author{\texorpdfstring{\large}{}Ruijie Gao\affmark{\ast}, Jirong Yang\affmark{\dagger},
  Mengtian Hong\affmark{\ddagger}, Youyou Lu\affmark{\ast}, Barry Lyu\affmark{\ast},
  Haoran Jin\affmark{\ast}, and Nathan Bleier\affmark{\ast}}
\affiliation{\institution{$^{\ast}$University of Michigan \quad
    $^{\dagger}$The University of Texas at Austin \quad
    $^{\ddagger}$University of Glasgow}
  \institution{$^{\ast}$\{ruijieg,yoyolu,barrylyu,allenjin,nbleier\}@umich.edu \quad
    $^{\dagger}$polarisyjr@utexas.edu \quad
    $^{\ddagger}$2960081H@student.gla.ac.uk}
  \country{}}
\renewcommand{\shortauthors}{Gao et al.}

\begin{abstract}

Emerging accelerators rarely ship with mature compiler backends, so recent work
asks LLM agents to write native kernels directly from ISA documentation and
simulator feedback. On two such accelerators, we find that this direct workflow
returns a correct kernel on only 17 of 38 workload instances, and that most of
its inference budget is spent re-deriving machine semantics and repairing
illegal code rather than optimizing the workload.

We present \systemname{}, which consists of two parts. Contour-Explore
replaces native-code generation with search over
a mapping surface: a typed set of scheduling knobs and their legal domains,
generated automatically from the accelerator's ISA description and the workload
graph. A compiler-owned resolver completes partial assignments into legal plans
and emits native code, so every evaluated candidate is verified against the
compiler's default. Kernel writing becomes knob assignment, and kernel
optimization becomes design-space exploration that any policy can drive through
the same intent interface, including hill climbing,
evolutionary search, and LLM agents. Compared with KernelCraft, a
state-of-the-art agentic kernel generator, an LLM agent through Contour-Explore
returns a verified kernel on every Gemmini and PLENA instance, produces
$2.86\times$ and $1.13\times$ faster kernels on the instances where both
succeed, and uses 26.3\% and 83.4\% fewer tokens per verified kernel.
KernelCraft returns a correct kernel on only 60\% of the Gemmini
instances and 5 of the 18 PLENA instances. The agent's kernels are also
$1.17\times$ faster than those from the expert-written tiling heuristics in
Gemmini's software library and $1.46\times$ faster than those from PLENA's
compiler, and
AccelOpt finds $1.50\times$ faster kernels on Gemmini and $1.20\times$
faster kernels on four PLENA attention instances when it searches through
Contour-Explore instead of editing kernel programs.

A generated surface is bounded by its vocabulary. The second component---Contour-Ext---lifts this bound: when
search stalls, an LLM agent reads the stalled program, hardware facts, and
search history, then extends the backend itself by implementing a new knob and
its lowering. Contour-Ext accepts the extension only if the
default path is byte-identical, every new value is numerically correct and
changes the program, and the enlarged surface yields a measured gain.
On VTA, Contour-Ext extends the AutoTVM surface of TVM, and AutoTVM's
XGBTuner then finds $4.14\times$ faster kernels; on PLENA, the extended
surface makes the kernels found by hill climbing $1.20\times$ faster.
 \end{abstract}

\maketitle

\section{Introduction}

Programmable tensor accelerators often expose target-specific
instruction set architectures, explicitly managed on-chip 
memories, and specialized data-movement and control mechanisms
\cite{chen2018isa4nn,moreau2019hardware,genc2021gemmini,jouppi2017tpu}.
{Supporting such a target requires a tensor compiler backend to lower tensor
operations to these interfaces, including instruction selection, memory
placement, state configuration, and legal instruction sequencing
\cite{hong2025autocomp,ikarashi2022exo}. However, most proposed tensor
accelerators lack reusable tensor compiler backends for generating
implementations beyond supported kernels or templates, and rapid hardware
iteration makes separately engineered backends difficult to maintain
\cite{act-oopsla2026,huang20243la,hong2024llmaidedcompiler,ikarashi2022exo}.}
When a reusable backend is unavailable, developers
often fall back on manually written low-level kernels or target-specific
compilation templates \cite{act-oopsla2026,nie2026kernelcraft}.

{Recent work has explored large language model (LLM)-based kernel generation
and optimization across GPUs and specialized tensor
accelerators \cite{hong2024llmaidedcompiler,zhang2025cudaforge,
  ouyang2025kernelbench}.
These systems construct and revise kernel programs, guided by compilation,
correctness tests, and performance feedback
\cite{nie2026kernelcraft,hong2025autocomp,zhang2025accelopt,
  sereda2025kforge,kernelevolve2026}.
For example, KernelCraft generates and repairs ISA-level kernels from
machine specifications, using compilation, execution, and correctness
feedback \cite{nie2026kernelcraft}, and AccelOpt starts from a correct
kernel and improves performance using beam search, profiling feedback, and
optimization memory distilled from slow--fast kernel pairs
\cite{zhang2025accelopt}. We refer to workflows in which agents generate or
revise complete kernel programs as \emph{program-level agentic generation}.}

\begin{figure}[t]
  \centering
\includegraphics[width=.97\columnwidth]{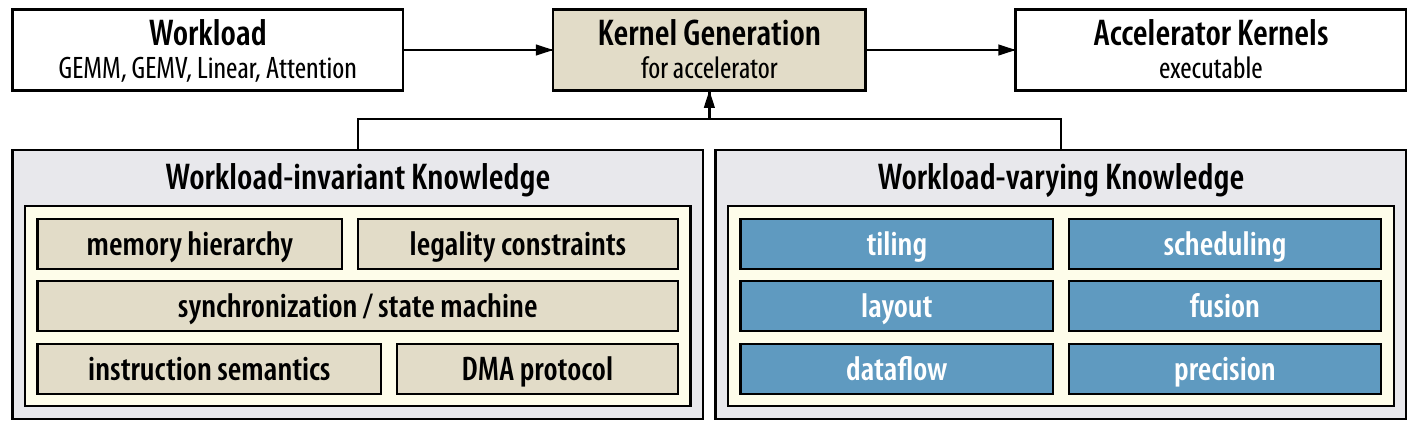}
  \caption{Kernel generation combines reusable machine-interface knowledge
  with workload-dependent mapping decisions.}
  \label{fig:two-knowledge}
\end{figure}

{Across workloads, these workflows involve two classes of information with different
reuse properties (Figure~\ref{fig:two-knowledge}).}
Workload-invariant machine knowledge defines the target interface, including
instruction semantics, the memory hierarchy, direct memory access (DMA)
protocols, configuration-state transitions, synchronization, and legality
constraints. Workload-dependent mapping decisions determine how a particular
computation uses that interface, including tiling, scheduling, layout, fusion,
dataflow, and precision. Both are required to produce an executable kernel, but
machine knowledge can be reused across workloads, whereas mapping decisions
generally vary across workload instances.
Without a reusable backend, the agent handles both within the same
per-workload loop, so each invocation spends inference and validation
feedback not only on workload-specific optimization, but also on instruction
legality, configuration-state transitions, data-movement protocols, and
synchronization.

Recent advances make this coupling optional. TensorLift recovers executable
tensor-level machine semantics from RTL
\cite{10.1145/3831252.3833946} and represents them in the Tensor Accelerator
ISA Definition Language (TAIDL) \cite{taidl-micro2025}, from which a
compiler backend can be generated automatically \cite{act-oopsla2026}.
Workload-invariant machine knowledge can therefore be compiled once and
reused across agent invocations.

\begin{figure}[t]
  \centering
  \includegraphics[width=\columnwidth]{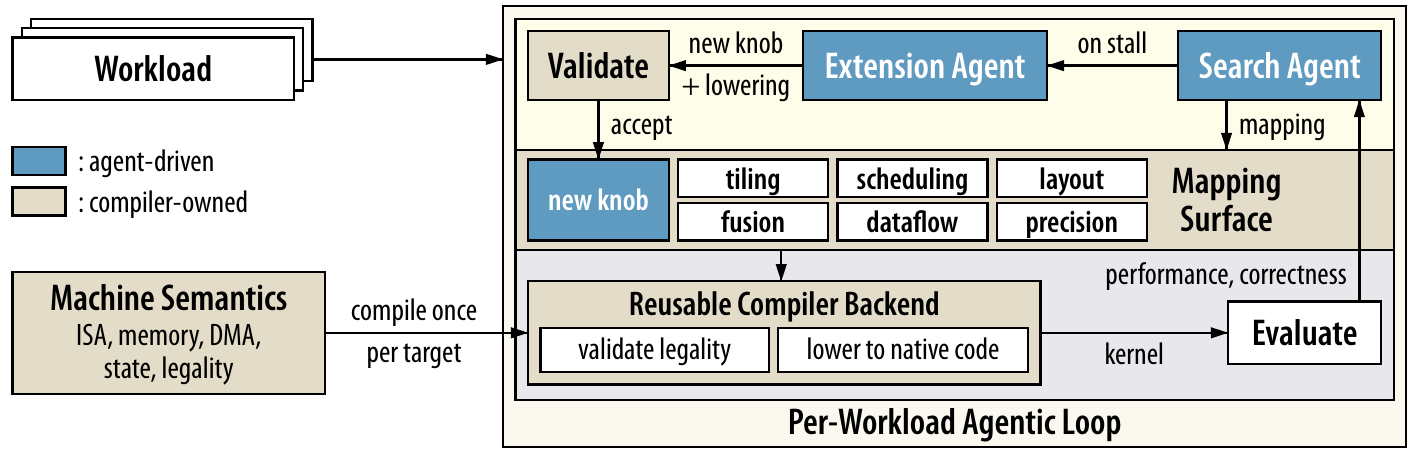}
\caption{Compiler-mediated agentic kernel generation compiles machine
  semantics once while searching workload-dependent mappings per
  workload, and extends the mapping surface with a validated knob when
  search stalls.}
  \label{fig:per-workload-agentic-loop}
\end{figure}

{A generated backend takes over machine-specific code generation. Two
problems remain. First, search agents need explicit control over mapping
decisions: where a compiler resolves implicit mapping choices,
program-level workflows treat that completion as a black box and lack
a shared contract for submitting partial mapping decisions and inspecting
their resolution.
Second, the compiler defines which mapping decisions an agent can
express, and this set bounds every search policy: a mapping that no
published decision expresses cannot be reached, whereas a program-level
agent can write it directly. We address the first problem by separating
mapping search from compiler-owned completion and native code generation,
and the second by letting an agent extend the compiler when search stalls
(Figure~\ref{fig:per-workload-agentic-loop}).}

{We present \systemname{}, an agentic framework in which search agents
choose mapping decisions and the compiler generates the kernel.
\systemname{} consists of two parts: \sysexplore{} searches a mapping
surface, and \sysext{} extends it.}

{\sysexplore{} derives a workload-conditioned mapping surface from
formal, RTL-lifted machine semantics and publishes each mapping decision as
a typed knob with its legal values. The agent submits a mapping intent:
required or preferred knob values, with other decisions left to the
compiler. For example, it may select an algorithm, prefer a tile size, and
leave loop organization unspecified. A symbolic resolver completes a legal
mapping for the backend to lower to native code; completing an empty
intent yields a verified default kernel, \candidatezero{}, which initializes
search. During search, feedback reports the realized mapping values,
correctness, and performance to guide the agent's next mapping hypothesis.
A \emph{search agent}, the component that proposes mapping intents,
can implement an LLM-based policy or a conventional search policy, such as
hill climbing, evolutionary search, or random sampling; all policies
share the same mapping surface, symbolic completion, code generation, and
candidate-validation path.}

\sysext{} extends the mapping surface when search in \sysexplore{}
stalls. An LLM-based \emph{extension agent} analyzes the stalled kernel and
the search history, then implements a new knob and its lowering in the
compiler. \sysext{} keeps the knob only if it passes correctness and
effectiveness checks, after which \sysexplore{} republishes the surface and
search continues. The agent's ability to write low-level code is thereby
applied to the compiler instead of to a single kernel: an accepted knob is
checked once and then remains available to later workloads and to every
search policy. \sysext{} requires only a knob registry, a lowering, legality
checks, and an evaluator from its host compiler, so it also applies to
existing compilers such as TVM \cite{chen2018tvm}.

This paper makes the following contributions:

\begingroup

\begin{itemize}
  \item We characterize how program-level agentic generation repeatedly
  reconstructs workload-invariant machine semantics on accelerators without
  a reusable kernel-generation backend: when we replicate KernelCraft
  on PLENA, repairing programs and reasoning about machine semantics take
  $4.1\times$ the completion tokens spent on optimizing correct programs.

  \item We design \sysexplore{}, a workload-conditioned mapping-search
  interface that publishes typed decisions derived from the hardware
  specification and the workload. It supports partial intents with
  required, preferred, and automatic decisions, completes them into legal
  plans, and reports the realized mapping values as feedback to guide
  subsequent search.

  \item We design \sysext{}, which extends the mapping surface when
  search stalls with knobs that pass correctness and effectiveness checks.
  On 16 VTA instances, \sysext{} extends the AutoTVM surface of TVM, and
  AutoTVM's XGBTuner finds $4.14\times$ faster kernels on the extended
  surface than on the original one. On 18 PLENA instances, the extended
  surface makes the kernels found by hill climbing $1.20\times$ faster.

\item We evaluate \sysexplore{} against KernelCraft, AccelOpt, and
  the software shipped with each accelerator: the tiled matrix-multiply
  routine in Gemmini's official software library and the compiler of PLENA.
  An LLM agent searching through
  \sysexplore{} returns a correct kernel on every Gemmini and PLENA
  instance, whereas KernelCraft fails on 21 of the 38; where both succeed,
  the agent's kernels are faster and cost fewer tokens. The agent's
  kernels are also $1.17\times$ faster than the Gemmini library routine and
  $1.46\times$ faster than those of the PLENA compiler. AccelOpt also finds
  faster kernels when it searches through \sysexplore{} than when it edits
  kernel programs.
\end{itemize}
\endgroup

 \section{Preliminaries}

\subsection{Emerging Accelerator Architectures}
\label{sec:prelim-arch}

We study three representative accelerator platforms: Gemmini
\cite{genc2021gemmini}, an established state-of-the-art tensor-accelerator
generator; PLENA \cite{wu2026combating}, a recently introduced open-source
research accelerator for long-context agentic LLM inference; and VTA
\cite{moreau2019hardware}, an open-source deep-learning accelerator. All
three expose
target-specific instructions and explicitly managed on-chip memories, so
mapping a tensor computation requires coordinating its computational
decomposition with data placement and movement.

\noindent\textbf{Gemmini.}
Gemmini is an open-source, full-stack DNN accelerator generator integrated with
RISC-V systems. Its central compute structure is a configurable spatial array
whose processing elements perform
multiply--accumulate operations using weight- or output-stationary dataflow.
The array reads operands from a banked, explicitly managed scratchpad and
writes partial and final results to a higher-precision accumulator. A Gemmini
mapping therefore determines how matrices are tiled, staged through the
scratchpad, accumulated, and executed under the selected dataflow.

\noindent\textbf{PLENA.}
PLENA is designed for long-context agentic LLM inference and organizes its ISA
around Matrix, Vector, Scalar, HBM, and Control instruction classes. Its Matrix
Unit uses an output-stationary flattened systolic array: operands stream along
a long reduction dimension while sub-arrays retain partial sums, which are
combined by a cross-array reduction.
PLENA separates its on-chip storage into a Vector SRAM for activations and
vector intermediates and a Matrix SRAM for weights and KV tensors. The latter
supports transposed and non-transposed reads without explicitly rearranging
the stored data, directly supporting computations such as $QK^\mathsf{T}$.
The Vector Unit complements matrix multiplication with elementwise,
reduction, and nonlinear operations needed by normalization and softmax.

PLENA composes these capabilities at tile granularity. Independent HBM
prefetch paths can fill the Matrix and Vector SRAMs while other units execute,
and fine-grained scheduling can interleave matrix multiplication, vector
reduction, scalar control, and data movement within a fused attention
pipeline. The architecture also supports mixed-precision MX formats across
its storage and compute paths. Its programmable mapping surface consequently
includes matrix and vector tiling, cross-array reduction, tensor orientation,
SRAM placement and residency, prefetching and buffering, numerical precision,
fusion, and overlap across execution units. Together, these are hardware-level mapping
decisions that the compilation approaches below must ultimately realize.

\noindent\textbf{VTA.}
The Versatile Tensor Accelerator (VTA)~\cite{moreau2019hardware} is the
open-source deep-learning accelerator of the TVM compiler
stack~\cite{chen2018tvm}. Its GEMM core multiplies int8 tiles held in
explicitly managed on-chip buffers and accumulates in int32. Its load,
compute, and store modules run concurrently and synchronize through
dependency queues, so data movement can overlap computation. TVM generates
VTA code from a schedule, which fixes the tiling, loop order, and on-chip
buffer placement of a tensor operator; its VTA backend maps the scheduled
loops to VTA load, GEMM, ALU, and store instructions.

\subsection{Machine Semantics and Generated Backends}

We use \emph{compiler backend} to mean the target-specific component
that translates a tensor computation graph into accelerator assembly. Prior
work provides the machine semantics that a backend needs in machine-readable
form. The Tensor Accelerator ISA Definition Language (TAIDL) describes
accelerator state, instruction semantics, and legality constraints using
tensor operations \cite{taidl-micro2025}. TensorLift generates TAIDL from
RTL: it lifts bit-accurate architectural-state updates into tensor-level
descriptions of computation, data movement, and control, and validates them
against the low-level model with SMT proofs and simulation
\cite{10.1145/3831252.3833946}. A compiler backend can then be generated
from the TAIDL specification \cite{act-oopsla2026}; on seven Gemmini
workloads, TensorLift with such a backend achieves a $1.014\times$
geometric-mean speedup over the official handwritten kernels
\cite{10.1145/3831252.3833946}.

{These systems provide reusable machine semantics and native code generation,
but do not expose workload-level mapping decisions to an external search
policy or accept partially specified mapping intents. \systemname{} introduces
a mapping-search interface that publishes the decisions available for each
workload and accepts required, preferred, and automatic choices.
Its symbolic resolver completes partial intents into legal plans
and reports their realized mapping values as feedback. The mapping surface,
intent interface, resolver, and feedback-directed search workflow are
introduced by this work.}

\begin{table}[t]
  \caption{Tokens generated by \kernelcraftname{} on PLENA.}
  \label{tab:kc-budget}
  \centering
  \small
  \setlength{\tabcolsep}{4pt}
  \renewcommand{\arraystretch}{1.12}
  \arrayrulecolor{taxrule}
  \color{black}
  \begin{tabular}{lrr}
    \toprule
    \rowcolor{taxhead}
    \tblhead{Token use} & \tblhead{Completion} & \tblhead{Reasoning} \\
    \midrule
    \rowcolor{taxopta}
    Repairing illegal or incorrect programs & 5.79M & 4.34M \\
    \rowcolor{taxoptb}
    Reasoning about machine semantics & 3.99M & 3.90M \\
    \rowcolor{taxopta}
    \textbf{Repairing or machine semantics} & \textbf{8.28M} & \textbf{6.75M} \\
    \rowcolor{taxoptb}
    \textbf{Optimizing a correct program} & \textbf{2.04M} & \textbf{1.42M} \\
    \rowcolor{taxopta}
    Ratio & $4.1\times$ & $4.8\times$ \\
    \bottomrule
  \end{tabular}
  \arrayrulecolor{black}
\end{table}

\subsection{{Program-Level Agentic Generation}}

KernelCraft studies a different response to the absence of a reusable
backend: a tool-using LLM agent directly generates and repairs a close-to-metal
kernel for each workload \cite{nie2026kernelcraft}. Each task supplies a
concrete workload instance, the target ISA and hardware
documentation, and memory-layout information. A platform-specific execution
harness gives the agent access to the target compiler, simulator, correctness
oracle, and documentation retrieval. Together, these inputs and services form
the agent's context. The required output is an ISA-level program that
explicitly realizes both the computation and its use of the target machine.

Generation proceeds as a multi-turn diagnosis-and-repair loop. In each
iteration, the agent produces a candidate assembly program, compiles and
executes it, and receives syntax diagnostics, functional differences, and
execution latency. When additional machine information is needed, the agent
queries the supplied ISA and hardware documentation through
\texttt{grep\_docs}. It uses this feedback first to repair correctness and then
to optimize performance, repeating the loop until its interaction budget is
exhausted.

Because the generated artifact is already close to the target ISA, the agent
is responsible for more than workload mapping. Within the same program, it
must select instructions, calculate addresses, place values in the memory
hierarchy, schedule data movement and computation, manage configuration state,
and expose parallelism. KernelCraft shows that frontier agents can sometimes
produce correct, performance-competitive kernels for unfamiliar accelerators.

\noindent{Program-level agents can also optimize an existing kernel. AccelOpt
organizes this process around three agents: a planner, executor, and
summarizer \cite{zhang2025accelopt}. In its workflow, the planner
proposes optimizations using profiling
feedback and accumulated experience; the executor implements these plans as
kernel programs; and the summarizer distills reusable optimization insights
from slow--fast kernel pairs for subsequent exploration. In its Trainium
implementation, the agents receive hardware constraints and NKI programming
guidance through platform-specific prompts.}

\subsection{Motivation}

{KernelCraft's program-level generation workflow treats each workload task in isolation.}
Although the target ISA, memory organization, and instruction semantics remain
unchanged across tasks, the workflow does not carry the machine-specific
knowledge recovered for one task into the next. Consequently, the target
details required for native-code generation remain part of every per-workload
interaction.

When we replicate KernelCraft's experiments on PLENA with its benchmark
workloads, including GEMM, GEMV, normalization, softmax, FFN, and attention,
and classify each model turn by tool verdict and by content,
most completion tokens go to repairing programs and to reasoning about
\emph{machine semantics}: instruction semantics, operand conventions, memory
layout, precision formats, legality, and timing. These two uses take
$4.1\times$ the completion tokens and $4.8\times$ the reasoning tokens spent
on optimizing correct programs (Table~\ref{tab:kc-budget}). Every run
spends 9.7--50.9\% of its completion tokens on this workload-invariant
knowledge.

This observation motivates our compiler-mediated agentic design: a reusable
compiler captures target-invariant code-generation knowledge and
generates every candidate from the validated machine semantics, so that
candidates are legal by construction and the agent can spend its tokens on 
workload-specific mapping decisions instead of program repair.
 \section{Design}
\label{sec:design}

\begingroup

\systemname{} consists of two parts (Figure~\ref{fig:overview}).
\textbf{\sysexplore{}} connects a search agent to a reusable compiler through a
workload-specific mapping interface. The agent chooses how to map a tensor
computation; the compiler completes these choices and generates target code.
An independent evaluator checks the output and measures performance. A
controller returns these results to the agent and retains the fastest verified
candidate. The same interface supports neural and conventional search policies.
\textbf{\sysext{}} extends this interface when search stalls. An extension
agent implements a new knob and its lowering in the compiler, and \sysext{}
keeps the new knob only if it passes correctness and effectiveness
checks.
Sections~\ref{sec:design-surface}--\ref{sec:design-codegen} describe
\sysexplore{}, and Section~\ref{sec:design-ext} describes \sysext{}.

\begin{figure}[t]
  \centering
  \includegraphics[width=\columnwidth]{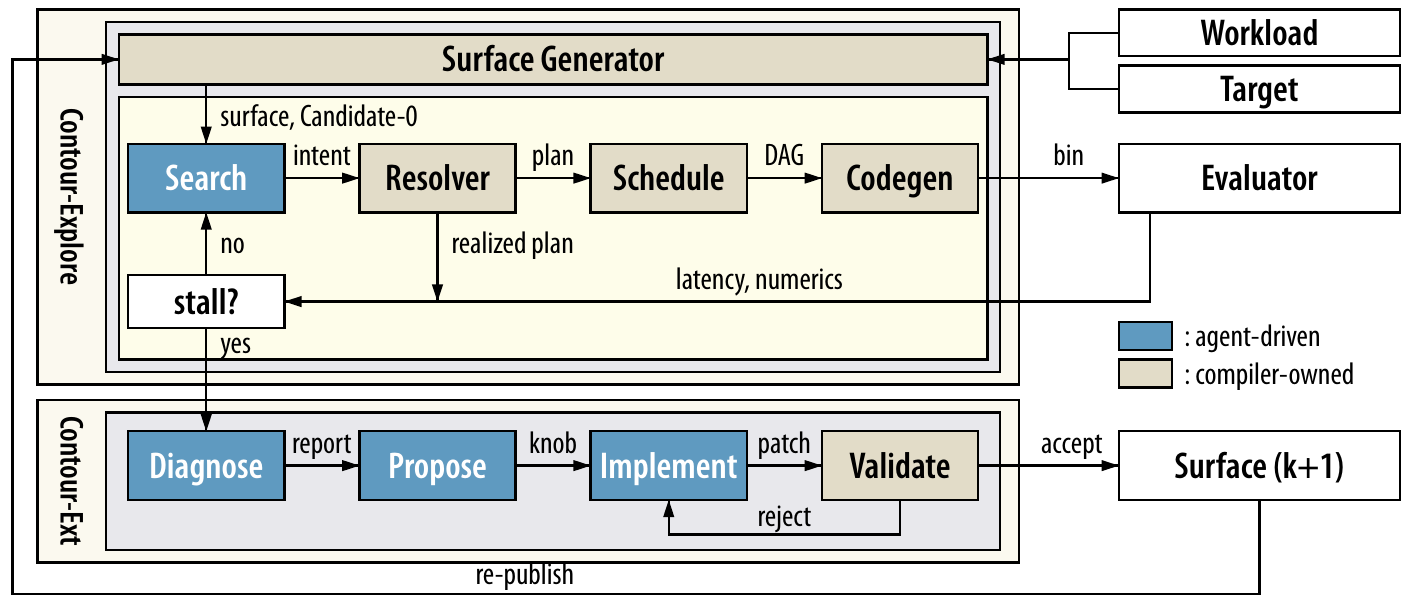}
\caption{\systemname{} overview: \sysexplore{} searches the published
  mapping surface, and \sysext{} extends it when search stalls.}
  \Description{Block diagram. In Contour-Explore, a surface generator reads the
  workload and target and publishes a surface and Candidate-0 to search.
  Search sends intents to the resolver, which passes plans to the schedule
  layer and code generation; the evaluator returns latency and numerics.
  When search stalls, Contour-Ext diagnoses, proposes a knob, implements a patch,
  and validates it; an accepted patch produces the next surface.}
  \label{fig:overview}
\end{figure}

\begin{algorithm}[!t]
  \color{black}
  \caption{{Mapping search with explicit preference completion}}
  \label{alg:mapping-search}
  \footnotesize
  \noindent\begin{tabular}{@{}lll@{}}
    \surfaceTag\surfaceHighlight{Surface} &
    \hardTag\hardHighlight{Hard checks} &
    \preferenceTag\preferenceHighlight{Preferences} \\
    \regionTag\regionHighlight{Region completion} &
    \executionTag\executionHighlight{Compile/evaluate} &
    \selectionTag\selectionHighlight{Selection}
  \end{tabular}\par\smallskip
  \begin{algorithmic}[1]
    \Require workload $W$, target semantics $T$, search budget
    \State \surfaceTag\surfaceHighlight{$S \gets \Call{DeriveSurface}{W,T}$}
    \State $p_0 \gets \Call{Resolve}{\varnothing,W,T}$
    \State \executionTag\executionHighlight{$b \gets \Call{CompileAndEvaluate}{p_0,W}$}
    \State Require $b$ to pass verification; $F \gets \{b\}$
    \While{proposal and evaluation budgets permit another batch}
      \State $I \gets \Call{Propose}{W,S,F}$
      \If{target is Gemmini}
        \State $p \gets$ catalog plan selected by $I$
        \State $P \gets \{p\}$ if $p$ is fresh, else $\varnothing$
      \Else
        \State Check region $I$ and \hardHighlight{completion of its hard subset}
        \State \regionTag\regionHighlight{Sample $I$; complete intents with \textsc{Resolve}}
        \State \regionHighlight{Deduplicate plans}; expand a short neighborhood
        \State \regionHighlight{Filter static-cost outliers; select two diverse plans}
        \State $P \gets$ selected plans, or $\varnothing$ on rejection
      \EndIf
      \ForAll{$p \in P$}
        \State \executionTag\executionHighlight{$r \gets \Call{CompileAndEvaluate}{p,W}$}
        \If{\selectionTag\selectionHighlight{$r.\mathit{verified}$ and $r$ is faster than $b$}}
          \State \selectionHighlight{$b \gets r$}
        \EndIf
      \EndFor
      \State $F \gets$ mapping outcomes, incumbent, remaining budget
    \EndWhile
    \State \Return $b$
    \Statex \hrulefill
    \Function{Resolve}{$I,W,T$}
      \State \hardTag\hardHighlight{$A \gets$ hard assignments in $I$}
      \State $p \gets$ complete $A$ with defaults; check joint constraints
      \State \hardHighlight{Reject if completion fails}
      \ForAll{preferences $(k,v)$ in \preferenceTag\preferenceHighlight{fixed order}}
        \State $q \gets$ complete and check $A \cup \{k=v\}$
        \If{$q$ is legal}
          \State \preferenceHighlight{$A \gets A \cup \{k=v\}$; $p \gets q$}
        \Else
          \State \preferenceHighlight{Record the relaxation and its reason}
        \EndIf
      \EndFor
      \State Construct plan sections and schedule DAG; reject cycles
      \State \Return plan with realized decisions and resolution records
    \EndFunction
  \end{algorithmic}
\end{algorithm}

\subsection{Constructing the Mapping Surface}
\label{sec:design-surface}

The input is a tensor computation in the high-level optimizer (HLO)
intermediate representation~\cite{openxla_stablehlo} and a target described by TAIDL. A
\emph{workload contract} fixes the computation, input data, target,
execution boundary, and numerical acceptance policy for every candidate.
Within this contract, a \surfaceTag\surfaceHighlight{\emph{mapping surface}} lists the decisions that search
may change, such as tiling, operand residency, or execution overlap. Each
decision, or \emph{knob}, has a type and a finite domain of supported values.

The surface generator constructs a workload-specific mapping surface using a TAIDL
hardware description and a shared mapping layer. TAIDL specifies architectural
state, storage capacities, instruction semantics, and legality constraints.
The shared mapping layer defines the knob vocabulary and default policies,
which are compiler design choices reused across accelerators. For each
workload, the surface generator uses the hardware description and these shared policies
to \surfaceHighlight{derive the available knob values}, and the resolver
checks their joint legality.

The shared mapping layer defines each knob category with four parts:
a derivation rule that creates knob instances and candidate domains from the
workload graph and TAIDL facts, a default policy, a legality predicate for
its values, and a plan field that lowering reads. We call these four parts a
\emph{knob definition}. A \emph{compiler version} is the compiler together
with its registered knob definitions. The surface generator publishes one
surface for each compiler version. \sysext{} (Section~\ref{sec:design-ext})
registers new knob definitions through the same interface, so added knobs
and built-in knobs use the same derivation, publication, and completion
steps.

On PLENA, the surface generator first selects the knob categories applicable to the
operator family. The eight categories cover algorithm, matrix and vector
decomposition, placement, memory pipeline, execution pipeline, precision, and
loop structure. For each applicable knob, it tests each supported value by
requiring that value and completing the remaining decisions with the default
policies. It retains values whose completion passes the constraints and omits
knobs with only one retained value. Those fixed choices remain compiler-owned.

These per-knob tests do not make the domains independently combinable. The
resolver checks joint constraints again when it completes an intent. For
example, the attention algorithm determines whether scores are materialized;
that choice changes SRAM demand and the legal block sizes. A conditional
surface therefore publishes individual domains without enumerating their
cross-product. Gemmini instead uses a finite catalog of complete mappings
that pass the workload's shape, capacity, and scheduling checks. Both forms
expose legal choices without exposing physical instructions or addresses.

\subsection{Completing Mapping Intents}

An \emph{intent} assigns values to exposed knobs. \systemname{} supports
\hardTag\hardHighlight{\emph{hard}} assignments that must hold and
\preferenceTag\preferenceHighlight{\emph{preferred}} assignments that may be relaxed.
The resolver rejects unknown knobs and values outside the published domains,
and fills automatic or omitted choices using shared default policies.
Gemmini selects a complete catalog mapping and treats all selected values
as hard assignments; PLENA also supports partial intents with preferences.

For PLENA, completion first applies all \hardHighlight{hard assignments} together. The
resolver fills the remaining choices with the shared default policies and
checks shape, capacity, reduction, output-coverage, and dependency constraints.
If this completion fails, it \hardHighlight{rejects the intent}. It then processes preferences
in a \preferenceHighlight{fixed order}: category order followed by knob-name order. For each
preference, it adds the requested assignment to the accepted assignments and
attempts completion again. A successful check \preferenceHighlight{retains the assignment}; a
failure \preferenceHighlight{drops it} and records the requested value, realized value, and reason.
Preferences are incorporated incrementally in a fixed order, retaining each
choice only when its completion remains legal alongside previously accepted
assignments.

Mapping completion produces a \emph{resolved plan}: a complete legal mapping with eight
semantic sections for the computation graph, matrix and vector decomposition,
placement, memory, precision, pipeline, and loops. It also constructs an
engine-schedule directed acyclic graph (DAG) specifying operations and
dependencies, rejecting cyclic schedules. Internal resolution records identify
defaulted choices and honored or relaxed preferences. The resolved decisions
guide subsequent lowering of the tensor computation to target code.

\subsection{Feedback-Directed Search}

Search begins by completing the empty intent. The resulting default plan,
\candidatezero{}, uses the shared default policies under the target and
workload constraints. It requires no model inference. The controller compiles,
executes, and verifies this plan before using it as the initial
\emph{incumbent}, the fastest verified candidate so far. The agent receives
the workload, surface, default decisions and performance, and remaining
evaluation budget. PLENA additionally reports a bottleneck profile containing
engine activity, HBM traffic, and SRAM occupancy.

\noindent\textbf{Gemmini: selecting a legal point.}
The agent sees the catalog's mapping decisions, with performance hidden for
untried mappings. It submits one complete mapping per evaluation. The
controller rejects duplicate submissions and returns verification, measured
cycles, and structural properties for each newly evaluated mapping. The agent
can then refine a tile choice or explore different buffering and store
schedules. Only candidates evaluated within the run enter final selection.

\noindent\textbf{PLENA: proposing a region.}
A \regionTag\regionHighlight{\emph{search region}} describes how the controller should generate candidate
intents. Hard assignments remain fixed; preferred assignments guide sampling;
explicitly automatic knobs are left to compiler completion. Other unassigned
knobs remain available for sampling. Each region includes at least one
\emph{diversity axis}, identifying a knob along which alternative plans are
desired, and leaves at least two knobs free of hard assignments. It may also
name an evaluated plan as a center and specify an initial radius in changed
decisions. Diversity guides sampling and selection without adding legality
constraints.

The controller constructs a center using preferred values, then the named
plan's remaining values, or domain midpoints when no plan is named. Explicitly
automatic knobs remain omitted from \regionHighlight{candidate intents}. Without a named
plan, the controller \regionHighlight{samples} up to 256 assignments: preferred knobs take
their requested values with probability 0.75 and otherwise draw uniformly;
other free knobs draw uniformly. With a named plan, sampling changes one to
the radius's number of knobs, prioritizing diversity axes. Each sampled
assignment becomes a candidate intent whose sampled values are preferences
and whose hard assignments are preserved from the region.

The controller \regionHighlight{removes rejected intents} and plans that duplicate either
previous evaluations or other plans in the pool. Deduplication uses the eight
semantic sections. If a neighborhood supplies fewer than two fresh plans, it
expands the radius up to eight; a pool still smaller than two is rejected.
It \regionHighlight{filters the worst quarter} by estimated static cost, preserving the center
and at least two plans. This estimate filters candidates rather than choosing
the measured winner.

\regionHighlight{Two plans enter evaluation}. The first is the center when available; otherwise,
it is the plan most distant from previously evaluated plans. The second
prioritizes a new value on the first diversity axis, then maximizes distance
from the first plan and evaluated plans. Distance is normalized Hamming
distance over the eight semantic sections. Both plans obey the hard
assignments, but their realized preferences can differ after completion.

Algorithm~\ref{alg:mapping-search} combines this target-specific proposal path
with the common search loop. Feedback reports each plan's identity,
verification, measured performance, and realized decisions relative to the
default. A rejected region instead receives a mapping-level explanation,
such as conflicting hard assignments or exhausted alternatives. The agent can
revise its proposal, but only the controller can update the incumbent.
The controller bounds both proposals and evaluations. Rejected regions consume
requests without consuming evaluations, so repeated rejection cannot prolong
search indefinitely. A new batch requires sufficient remaining evaluation
budget; \candidatezero{} is outside this search budget.

\noindent{\textbf{Conventional search policies.}
Hill climbing, genetic search, and random sampling implement the
\textsc{Propose} step in Algorithm~\ref{alg:mapping-search} over the published
mapping decisions. Hill climbing explores neighboring assignments, genetic
search mutates and recombines typed intents, and random search samples
assignments from the exposed domains. \systemname{} completes these proposals
into legal mappings and generates kernels for verification and measurement.
Measured latency provides the objective value for hill climbing and genetic
selection; random sampling proceeds independently of these values. Across
all policies, the controller retains the fastest verified candidate. Thus,
conventional policies search mapping choices without implementing kernel
generation or legality checks.}

\noindent\textbf{Stall detection.}
Search \emph{stalls} when the incumbent stops improving: the relative
latency improvement over a window of recent evaluations falls below a
threshold, or the policy can reach no unevaluated program. An exploration
round also ends when an LLM search agent produces no new evaluation for a
fixed number of turns, or when the round's evaluation budget is spent. The
incumbent at the end of the round is the \emph{stalled plan}, $p_s$.
\sysext{} (Section~\ref{sec:design-ext}) starts from $p_s$.

\subsection{Code Generation and Candidate Validation}
\label{sec:design-codegen}

\noindent\textbf{Gemmini realization.}
\systemname{} exposes macro tiling, buffering, store scheduling, and batch-region
choices through a legal mapping catalog. Macro-tile dimensions are multiples
of the primitive tile, bounded by workload dimensions and scratchpad and
accumulator capacities. The selected mapping specifies the choices that the
subsequent schedule transformations must realize; choices with only one
applicable value remain fixed.

The existing generated backend \executionTag\executionHighlight{lowers the HLO computation}
to a primitive schedule with concrete transfer and matrix-operation addresses
and sizes. \systemname{} then applies schedule transformations to implement
accumulator initialization, the selected buffering strategy, and store order.
The schedule parser checks the transformed schedule before code generation.
This additional transformation stage connects the searchable mapping choices
to their executable realization.

\noindent\textbf{PLENA realization.}
The resolver translates the completed mapping into typed bridge options that
\systemname{} passes to code generation, making the resolved computation,
placement, data movement, and execution overlap explicit inputs to lowering.
Instruction selection, physical memory management, and synchronization realize
these decisions as \executionHighlight{target code}. From the same options,
the resolver derives an engine-schedule DAG for symbolic legality checks and
cost estimation. The mapping interface thus gives search control over the
plan realized by the backend.
A schedule layer reads the plan fields and selects operator
decompositions, placement, tensor storage, and loop nests before instruction
selection. Instruction selection maps abstract tile operations to target
instructions by semantic signatures derived from TAIDL. Lowering either
implements every value in a resolved plan or reports that it cannot; it
does not fall back to another value. On request, lowering tags each
instruction with the schedule-layer step that produced it.

\noindent\textbf{Independent validation.}
The evaluator \executionHighlight{executes} each generated program and
\executionHighlight{compares} its complete output with an independent numerical oracle under the workload contract.
The acceptance policy remains fixed across mapping and precision choices.
The evaluator returns correctness and performance to the controller,
together with structural counters when available. A candidate replaces the incumbent only when it
\selectionTag\selectionHighlight{passes verification} and \selectionHighlight{strictly improves} the performance objective. Otherwise,
the previous verified candidate remains the result. The agent cannot change
the acceptance criteria or the final selection rule.
The evaluator identifies each program by a hash of its assembly and
does not execute the same program twice. \systemname{} keeps the hashes of
all programs measured in the run.

\subsection{Extending the Mapping Surface}
\label{sec:design-ext}

\begin{algorithm}[!t]
  \color{black}
  \caption{Extending the mapping surface when search stalls}
  \label{alg:ext}
  \footnotesize
  \noindent\begin{tabular}{@{}lll@{}}
    \extDiagTag Diagnosis &
    \extProposeTag Proposal &
    \extImplTag Implementation \\
    \extStaticTag Correctness checks &
    \extSearchTag Effectiveness checks &
  \end{tabular}\par\smallskip
  \begin{algorithmic}[1]
    \Require workload $W$, target semantics $T$, compiler $C$, default plan
      $p_0$, stalled plan $p_s$, earlier rejections $H$
    \State \extDiagTag Profile $p_s$ and report its bottleneck $B$
    \While{the proposal budget permits}
      \State \extProposeTag $\kappa \gets \Call{ProposeKnob}{B,H}$
        \Comment{values and lowering}
      \State \textbf{if} no knob is proposed \textbf{then break}
      \State \extImplTag $C' \gets C$ with $\kappa$ registered and its
        lowering implemented
      \State $S' \gets \Call{DeriveSurface}{W,T,C'}$
      \State \extStaticTag Reject $\kappa$ if any of:
      \Statex \hspace{3.2em}$S'$ exposes no new knob
      \Statex \hspace{3.2em}a new value leaves the program unchanged
      \Statex \hspace{3.2em}the default value changes the program of $p_0$ or $p_s$
      \Statex \hspace{3.2em}a new value exceeds the output-error bound
      \State \extSearchTag $p^{*} \gets$ hill climbing on $S'$ from $p_s$
      \State \extSearchTag Reject $\kappa$ if any of:
      \Statex \hspace{3.2em}$p^{*}$ does not use $\kappa$
      \Statex \hspace{3.2em}the program of $p^{*}$ is reachable without $\kappa$
      \Statex \hspace{3.2em}$p^{*}$ is not faster than $p_s$ by the gain threshold
      \Statex \hspace{3.2em}a new value exceeds the error bound on scaled or
      \Statex \hspace{3.2em}\quad outlier inputs
      \State \textbf{if} $\kappa$ is not rejected \textbf{then return}
        $(C', p^{*})$
      \State $H \gets H \cup \{(\kappa, \text{rejection reason})\}$
        \Comment{revise or propose anew}
    \EndWhile
    \State \Return $(C, p_s)$
  \end{algorithmic}
\end{algorithm}

\sysext{} adds a knob to the surface when \sysexplore{} stalls. It uses four
parts of \sysexplore{}: the knob registry, the lowering, the legality checks,
and the evaluator. Algorithm~\ref{alg:ext} shows one invocation.

\noindent\textbf{Analysis of the stalled plan.}
\sysexplore{} passes \sysext{} the stalled plan $p_s$, \candidatezero{}
($p_0$), the search history, the hashes of measured programs, and earlier
rejected knobs ($H$). In \extDiagTag diagnosis, an LLM-based extension
agent examines $p_s$ without editing code. It reads the assembly, with each
instruction tagged by the lowering step that produced it, and an emulator
profile of memory traffic and unit occupancy. It ends with a bottleneck
report $B$ that cites measurements and code locations.

\noindent\textbf{Implementation of a new knob.}
In \extProposeTag proposal, the agent specifies one knob that addresses $B$
and differs from every entry in $H$: its values with the default first, the
lowering change for each value, and the operator families it applies to.
In \extImplTag implementation, the agent registers the knob with a
derivation rule and a legality predicate, and implements its values in the
schedule layer. It cannot change the computation, instruction selection, or
the evaluator. Before submission, it can compile and emulate plans through
the \sysexplore{} resolver and evaluator.

\noindent\textbf{Correctness and effectiveness.}
\sysext{} accepts a knob only if it passes two groups of checks. The
\extStaticTag correctness checks require that the default value reproduces
the programs of the previous compiler version at $p_0$ and $p_s$, that each
new value changes the program, and that each new value keeps the output
error within the numerical bound of the target. The \extSearchTag
effectiveness checks run hill climbing on the extended surface $S'$ from
$p_s$. Its best plan $p^{*}$ must use the new knob, its program must not be
reachable without the knob, and it must be faster than $p_s$ by a gain
threshold. The new values must also meet the numerical bound at $p^{*}$ on
inputs with large magnitudes and outliers. A failed check returns its
reason to the agent, which revises the patch or proposes another knob; each
rejected knob is added to $H$.

\noindent\textbf{Return to \sysexplore{}.}
An accepted knob becomes part of a new compiler version. \sysexplore{}
publishes the extended surface and resumes search from $p^{*}$; its search
policies use the new knob in the same way as any other knob. When \sysext{}
returns no knob, \sysexplore{} continues with the previous compiler version.

\par
\endgroup
 \section{Evaluation}

\subsection{Setup}
\label{sec:eval-setup}

We use Gemmini to evaluate \sysexplore{}
(Section~\ref{sec:eval-gemmini}), VTA with TVM and AutoTVM to evaluate
\sysext{} (Section~\ref{sec:eval-vta}), and PLENA to evaluate the full
system (Section~\ref{sec:eval-plena}).

\noindent\textbf{Accelerator platforms.}
We evaluate \systemname{} on Gemmini, VTA, and PLENA. For Gemmini, we use the
\texttt{GemminiRocketConfig} provided by Chipyard
\cite{chipyard,genc2021gemmini}, whose default Gemmini configuration sets
\texttt{DIM=16} and instantiates a $16\times16$ systolic array. We build the
bitstream of this SoC with Vivado~2022.2, generate its software image with
PetaLinux~2022.2, and deploy the resulting system on a Xilinx ZC706 FPGA
evaluation board. We obtain Gemmini performance measurements from this
physical implementation.

For VTA, we use the TVM Versatile Tensor Accelerator
\cite{moreau2019hardware} and measure cycles with tsim, which simulates its
Chisel RTL with Verilator. The evaluator compares each output element by
element with a CPU reference.

{For PLENA, we use the authors' official RTL release and the transaction-level
mode of their official simulator \cite{wu2026combating,plena-simulator}. The
simulator executes compiled assembly against an explicit memory layout and
produces both numerical outputs and simulated latency.}

\noindent {\textbf{Models and workloads.}
All main experiments use DeepSeek-V4.1-Flash
\cite{deepseek2026deepseekv4}, with the same model settings across
the compared LLM methods; the extension agent of \sysext{} uses the same
model.
On Gemmini and PLENA, we adopt the workload
definitions and input shapes from KernelCraft~\cite{nie2026kernelcraft}.
Each workload--input-shape combination is a \emph{workload instance}.
Within each workload family, C1, C3, and C5 denote the smallest,
middle-sized, and largest workloads, respectively.
The evaluation covers 20 Gemmini instances and
18 PLENA instances: FFN, SwiGLU, FlashAttention, MHA, GQA,
and MQA, each with shapes C1, C3, and C5. These multi-stage workloads
exercise fusion and attention computations beyond standalone GEMM and GEMV.}
On VTA, we use 16 int8 instances with int32 accumulation, shift, and
clipping: FFN (two dense layers with ReLU), GEMM, and GEMV, each with shapes
C1--C5 from smallest to largest, and the fully connected layer of ResNet-18
(R18). AutoTVM~\cite{chen2018learning} tunes the knobs of a hand-written
TVM schedule template. The TVM~0.12 VTA dense template exposes tile sizes
for the batch, input-channel, and output-channel dimensions and the number
of virtual threads over output channels; we call the mappings expressible by
these knobs the \emph{AutoTVM surface}. The \emph{\sysext{} surface} of an
instance adds the knobs that \sysext{} accepts on it. Hill climbing runs on
the AutoTVM surface until it stalls; \sysext{} then extends the surface from
that stall, and hill climbing continues on the \sysext{} surface. We also run
the AutoTVM XGBTuner, GATuner, and RandomTuner on both surfaces with up to
512 measurements per run and report the median of three seeds.

\noindent {\textbf{Search agents and baselines.}
Our \emph{naive agent} follows the feedback-directed loop in Section~\ref{sec:design}:
an LLM proposes a mapping intent, receives correctness and performance
feedback, and uses that feedback to propose the next intent. We denote
this configuration Naive + \sysexplore{}. \sysexplore{}
retains the fastest verified kernel found during the run.
We also connect hill climbing and a genetic algorithm (GA) to the same
interface. These policies share the mapping surface,
symbolic completion, code generation, and evaluator.}

{\kernelcraftname{} generates kernels without a supplied starting
kernel and provides the ISA-level generation baseline. Section~\ref{sec:disc-accelopt} evaluates the AccelOpt optimization harness through \sysexplore{} and at the program level.
We also report the software shipped with each accelerator: Gemmini's
library matrix-multiply routine (\texttt{tiled\_matmul\_auto} in
\texttt{gemmini.h}) \cite{genc2021gemmini}, which we call the Gemmini
library, and the PLENA compiler \cite{plena-simulator}, which compiles each
instance once without search.}

\noindent{\textbf{Search budgets and reporting.}
\sysexplore{} search permits up to 80 candidate evaluations per PLENA
instance and 12 unique candidate evaluations per Gemmini instance.
On PLENA, hill climbing and GA use all 80 evaluations. The naive agent
can stop earlier: its run is also bounded in the number of LLM turns and
requests, and those that repair a rejected intent or yield no new candidate
consume this bound without an evaluation. We report the evaluation counts.
Methods with three runs report the median
latency for each instance. Token usage and evaluation counts are averaged
across the runs. The naive agent uses three runs per instance. On PLENA, GA uses three runs and hill climbing uses one; on Gemmini, hill climbing and GA use 1,000 seeds per instance. \kernelcraftname{} uses three runs per instance.}
To extend the PLENA surfaces, hill climbing runs in iterations of 16
evaluations and invokes \sysext{} when the relative improvement over the last
six iterations is below 0.2\%. The effectiveness checks run 40 evaluations of
hill climbing and require a gain of at least 1\%. The extension of an
instance ends after three consecutive rejected invocations, eight accepted
knobs, 4,000 evaluations, or 48 hours.

\noindent {\textbf{Metrics.}
We report geometric-mean speedup over a reference kernel, which is the
Gemmini library on Gemmini and \candidatezero{} on VTA and PLENA, and verified
coverage, the fraction of instances for which a method returns a verified
kernel. \emph{Tokens per verified kernel} equal the total prompt and
completion tokens across all runs divided by the number of runs that
return a verified kernel. Failed runs contribute their token usage to
the numerator. Each run contributes at most one kernel to the denominator.}

{We define \emph{disappointment} on an instance as
$d=1-t_{\mathrm{best}}/t_{\mathrm{method}}$, where
$t_{\mathrm{method}}$ is the latency returned by the method and
$t_{\mathrm{best}}$ is the best latency known for the instance. On
Gemmini, it is the lowest latency that the naive agent, hill climbing, or GA
finds in any run, so a method outside these three can have negative
disappointment; on PLENA, it is the lowest latency among the compared
methods and the longer search runs on the merged \sysext{} surface
(Section~\ref{sec:eval-plena}).
A method that returns no verified kernel has $d=100\%$.
We report mean disappointment across instances.}

\subsection{\sysexplore{} on Gemmini}
\label{sec:eval-gemmini}

\begin{figure}[t]
  \centering
  \includegraphics[width=\columnwidth]{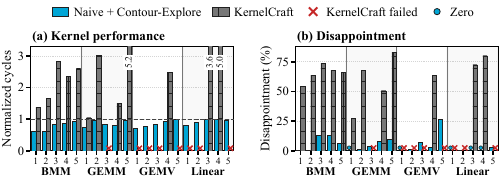}
  \par
  \setlength{\abovecaptionskip}{2pt}
\caption{Gemmini results on 20 instances: (a) cycles normalized to the
  Gemmini library (dashed line); (b) disappointment. Crosses:
  \kernelcraftname{} failures. BMM: BatchMatMul.}
  \label{fig:gemmini-search}
\end{figure}

\noindent{\textbf{Comparison.}
We compare the kernels found by the naive agent, hill climbing, and GA
through \sysexplore{} with the Gemmini library and with \kernelcraftname{}
across 20 instances from BatchMatMul, GEMM, GEMV, and Linear. Every kernel
uses the same workload inputs,
correctness oracle, and execution boundary. The naive agent, hill
climbing, GA, and \kernelcraftname{} have the same evaluation budget of 12
evaluations per instance. We also compare verified coverage and tokens per
verified kernel against \kernelcraftname{}.}

\noindent{\textbf{Search through \sysexplore{} against the hand-written library.}
Figure~\ref{fig:gemmini-search}a normalizes cycles to the Gemmini library,
the kernel that a Gemmini user obtains without search.
With 12 evaluations, the naive agent returns a faster kernel than the
library on 19 of the 20 instances, and its kernels are $1.17\times$ faster
in geometric mean. The gain holds in every workload family: $1.30\times$ on
BatchMatMul, $1.18\times$ on GEMV, $1.15\times$ on GEMM, and $1.06\times$ on
Linear, where the library is 0.4\% faster on Linear C4. Under the same
budget, hill climbing and GA are $1.17\times$ and $1.10\times$ faster than
the library. Mean disappointment is 4.9\% for both the naive agent and hill
climbing, 10.7\% for GA, and 18.2\% for the library: every policy is closer
on average than the library to the best kernel found on the surface.}

\noindent{\textbf{Verified coverage and kernel quality.}
\kernelcraftname{} returns a verified kernel on 12 of the 20 instances
(60\%). On the other eight, none of the programs that it generates within
its budget compiles and passes the correctness check; we count these
instances as failures. The naive agent through \sysexplore{} returns a
verified kernel on all 20 instances in each run.
On the 12 instances with a verified \kernelcraftname{} result, the naive
agent's kernels are $2.86\times$ faster than \kernelcraftname{}'s in
geometric mean, and \kernelcraftname{}'s kernels are slower than the
Gemmini library on all 12, by $2.44\times$ in geometric mean.
This performance comparison is conditioned on \kernelcraftname{} success;
the coverage comparison includes the entire suite. 
The naive agent returns a faster kernel than \kernelcraftname{} on every instance; mean disappointment is 78.6\% for \kernelcraftname{}, including its eight failures at 100\% (Figure~\ref{fig:gemmini-search}b).}

\noindent{\textbf{Cost per verified kernel.}
Table~\ref{tab:verified-token-cost} reports 564K tokens per verified
kernel for the naive agent through \sysexplore{}, compared with 765K for
\kernelcraftname{}, a 26.3\% reduction. The denominator accounts for the
\kernelcraftname{} runs that consume tokens without returning a
verified kernel. Coverage from the compiler default and verified incumbent
selection converts search expenditure into a usable result on every instance.}

\subsection{\sysext{} on VTA}
\label{sec:eval-vta}

\begin{figure}[t]
  \centering
  \includegraphics[width=\columnwidth]{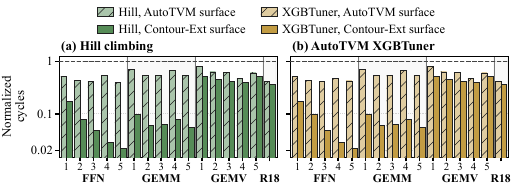}
  \par
  \setlength{\abovecaptionskip}{2pt}
\caption{VTA results on 16 instances: cycles normalized to
  \candidatezero{} (log scale). XGBTuner reports the median of three seeds.}
  \Description{Two side-by-side bar charts over 16 VTA instances with a
  logarithmic y-axis, one for hill climbing and one for XGBTuner. In both,
  bars on the Contour-Ext surface are far
  below the bars on the AutoTVM surface on FFN and GEMM, and slightly below
  them on GEMV and the ResNet-18 layer.}
  \label{fig:vta-ext}
\end{figure}

Hill climbing on the AutoTVM surface stalls at a $1.84\times$ geometric-mean
speedup over \candidatezero{}, which on VTA is the kernel compiled from the
default configuration of the TVM template. All 32 \sysext{} invocations, two per
instance, are accepted; they use 3.60M tokens in total, or 112K tokens per
invocation. On the resulting \sysext{} surface, hill climbing reaches
$7.77\times$, and its final cycles are $4.22\times$ lower than at the stall
in geometric mean (Figure~\ref{fig:vta-ext}a). The largest speedups occur on
FFN and GEMM, at $18.83\times$ and $14.10\times$ over \candidatezero{}; GEMV
and R18 reach $2.17\times$ and $2.78\times$. The most frequent added knobs
move operand loads out of inner loops, followed by knobs for loop order and
tiling. The added knobs also improve search procedures that were not used to
find them. XGBTuner reaches $1.85\times$ over \candidatezero{} on the AutoTVM
surface and $7.66\times$ on the \sysext{} surface, so its kernels are
$4.14\times$ faster on the \sysext{} surface
(Figure~\ref{fig:vta-ext}b), and GATuner and RandomTuner also reach fewer
cycles on the \sysext{} surface on all 16 instances. On FFN, most tuner
measurements on the AutoTVM surface fail because many tile combinations are
illegal; for XGBTuner with seed~1, the failure rate is 99.6\% on FFN C4.

\subsection{Full System on PLENA}
\label{sec:eval-plena}

On PLENA, we first measure how far \sysext{} raises the performance that
search can reach on the surface, and then evaluate \sysexplore{} on the
final surface against \kernelcraftname{}.

\begin{figure}[t]
  \centering
  \includegraphics[width=\columnwidth]{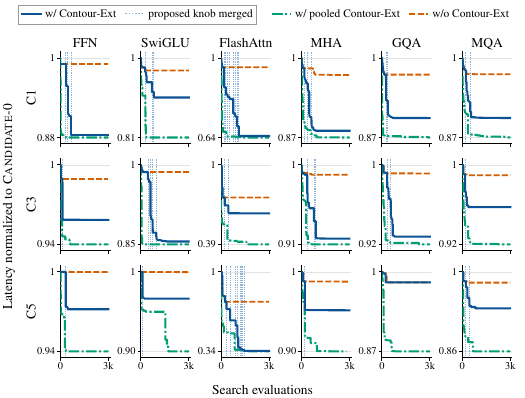}
  \par
  \setlength{\abovecaptionskip}{2pt}
\caption{Best latency so far on 18 PLENA instances, normalized to
  \candidatezero{}: base surface (w/o \sysext{}), per-instance \sysext{}
  surface (w/ \sysext{}; dotted lines mark accepted knobs), and merged
  \sysext{} surface (pooled).}
  \Description{Eighteen small step plots in a grid of three shapes by six
  workload families. In each plot, the curve on the base surface stays near
  one, the curve on the per-instance Contour-Ext surface drops after accepted
  knobs, and the curve on the merged surface drops early and ends lowest on
  most instances.}
  \label{fig:plena-ext}
\end{figure}

\noindent\textbf{\sysext{} raises the reachable performance.}
We call the surface that the surface generator publishes for an instance its
\emph{base surface}. On the base surface, hill climbing reaches a
$1.048\times$ geometric-mean speedup over \candidatezero{} after 3,000
evaluations per instance, and it improves \candidatezero{} by less than 3\%
on 14 of the 18 instances (Figure~\ref{fig:plena-ext}). \sysext{} is
invoked 119 times across the 18 instances and accepts 47 knobs, using 2.30B
tokens in total, or 19.3M tokens per invocation. With the same 3,000
evaluations, hill climbing on each instance's \sysext{} surface reaches
$1.191\times$. FlashAttention gains the most, from $1.229\times$ on the base
surface to $1.935\times$.

A knob definition derives its values from the workload, so a knob accepted
on one instance can be published on the others. We merge the knobs accepted
on the 18 instances into one surface with 30 knob definitions after
removing duplicates; \candidatezero{} is the same program on all surfaces.
On this \emph{merged \sysext{} surface}, hill climbing reaches $1.262\times$
after 3,000 evaluations (median of three seeds), which is
$1.20\times$ faster than its result on the base surface, and $1.157\times$ after the
first 80 evaluations. The merged surface reaches at least the speedup of the
instance's own \sysext{} surface on 17 of the 18 instances. On GQA C5, where
\sysext{} accepted no knob, it raises the speedup from $1.017\times$ to
$1.135\times$.

\begin{figure}[t]
  \centering
  \includegraphics[width=\columnwidth]{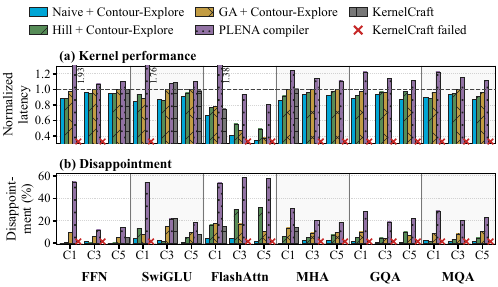}
  \par
  \setlength{\abovecaptionskip}{2pt}
\caption{PLENA results on the merged \sysext{} surface (18 instances):
  (a) latency normalized to \candidatezero{} (dashed line); (b)
  disappointment. Crosses: \kernelcraftname{} failures. Hill: first 80
  evaluations of one run.}
  \Description{Two bar charts over 18 PLENA instances for the naive agent,
  hill climbing, GA, the PLENA compiler, and KernelCraft. The naive agent has
  the lowest latency on most instances and disappointment near zero; the PLENA
  compiler is slower than Candidate-0 on 16 instances; KernelCraft
  fails on 13 instances.}
  \label{fig:plena-search}
\end{figure}

\noindent\textbf{\sysexplore{} on the merged surface.}
We run the naive agent, hill climbing, and GA through
\sysexplore{} on the merged \sysext{} surface with the 80-evaluation budget and
compare them with \kernelcraftname{} and the PLENA compiler
(Figure~\ref{fig:plena-search},
Table~\ref{tab:plena-policies}). Disappointment is measured against the best known
latency on the merged surface, found by the longer runs in
Figure~\ref{fig:plena-ext}, which reach $1.269\times$ speed-up over
\candidatezero{}.

\noindent\textbf{Search efficiency.}
The naive agent improves \candidatezero{} on all 18 instances and reaches a
$1.243\times$ geometric-mean speedup with a mean of 67.9 evaluations per
instance. Its mean disappointment is 2.1\%, and its disappointment
is below 3\% on 15
instances; on MHA C1, it finds a kernel 0.4\% faster than the best
one from the longer runs. Hill climbing and GA reach $1.157\times$ and
$1.139\times$, with mean disappointment of 8.3\% and 10.1\%; the naive
agent's kernels are 7.4\% and 9.1\% faster than theirs in geometric mean.
Among the three policies, the naive agent returns the lowest latency
on 15 instances. The largest gaps occur on FlashAttention C3 and C5, where
hill climbing has 30.2\% and 31.9\% disappointment after 80 evaluations and the
naive agent has 4.1\% and 1.6\%.

\begin{table}[t]
  \centering
\caption{\sysexplore{} search policies on the merged \sysext{} surface (18
  PLENA instances). Evaluations: mean per instance, with a budget of 80.}
  \label{tab:plena-policies}
  \begingroup\small
  \setlength{\tabcolsep}{5pt}
  \begin{tabular}{lrrr}
    \toprule
Policy & Speedup & Disappointment & Evaluations \\
    \midrule
    Naive & $1.243\times$ & 2.1\% & 67.9 \\
    Hill & $1.157\times$ & 8.3\% & 80.0 \\
GA & $1.139\times$ & 10.1\% & 80.0 \\
\bottomrule
  \end{tabular}
  \endgroup
\end{table}

\noindent\textbf{Comparison with \kernelcraftname{}.}
\kernelcraftname{} returns a verified kernel on 5 of the 18 instances and
fails on the other 13 by the criterion of Section~\ref{sec:eval-gemmini},
while every policy through \sysexplore{} returns one on all 18. On these five
instances, \kernelcraftname{} reaches $1.045\times$ over \candidatezero{}
with 13.0\% mean disappointment, and the naive agent reaches $1.181\times$ with
1.8\%; the naive agent is faster on all five. Pooling tokens over all runs,
the naive agent uses 1.393M tokens per verified kernel and
\kernelcraftname{} uses 8.375M, an 83.4\% reduction
(Table~\ref{tab:verified-token-cost}).

\noindent\textbf{Comparison with the PLENA compiler.}
The speedups above are measured over \candidatezero{}, and \candidatezero{}
is itself faster than the PLENA compiler on 16 of the 18 instances: the
compiler's programs reach $0.849\times$ the speed of \candidatezero{} in
geometric mean. The two exceptions are FlashAttention C3 and C5, where the
compiler skips key blocks that the causal mask excludes entirely and
\candidatezero{} computes them. Every policy through \sysexplore{} returns a
faster kernel than the compiler on all 18 instances, including these two,
and the naive agent's kernels are $1.46\times$ faster in geometric mean
(Figure~\ref{fig:plena-search}).

\begin{table}[t]
  \centering
\caption{Tokens per verified kernel. Both methods use three runs per
  instance.}
  \label{tab:verified-token-cost}
  \begingroup\small
  \setlength{\tabcolsep}{4pt}
  \begin{tabular}{lrrr}
    \toprule
Platform & Naive + \sysexplore{} & KernelCraft & Reduction \\
    \midrule
    Gemmini (20) & 564K & 765K & 26.3\% \\
    PLENA (18) & 1.393M & 8.375M & 83.4\% \\
    \bottomrule
  \end{tabular}
  \endgroup
\end{table}

 \section{Discussion}

\subsection{Sensitivity to Model Capability and Optimization Harness}
\label{sec:disc-accelopt}

\begingroup

We examine whether changing the model improves program-level generation
and search through \sysexplore{}. The Gemmini ablation compares Flash and Pro
within the DeepSeek-V4 family; the PLENA experiments extend the comparison
to GPT-6.1-Sol (Sol), separating generation success
from the performance of verified kernels.

\noindent\textbf{Gemmini: limited gains from Flash to Pro.}
We retain six instances selected from the Flash results to span easy,
intermediate, and success-boundary cases. Each model--method pair reports
seed~1, with candidates replayed under the same board timing harness
(Figure~\ref{fig:model-ablation}). For Naive + \sysexplore{}, five instances
differ by at most 0.1\%; Pro improves GEMM C4 by 4.7\%. Across all six,
Pro lowers the geometric-mean cycle count by 0.8\% while increasing API cost
by $17.0\times$. KernelCraft's response is mixed: Pro solves GEMV C3,
where Flash fails, but fails on Linear C4, which Flash solves. Among the five
instances solved by KernelCraft-Pro, Naive + \sysexplore{} + Flash is
$1.60\times$ faster in geometric mean. These results establish limited
benefit from this model upgrade on the tested Gemmini mappings.

\begin{figure}[t]
  \centering
  \includegraphics[width=\columnwidth]{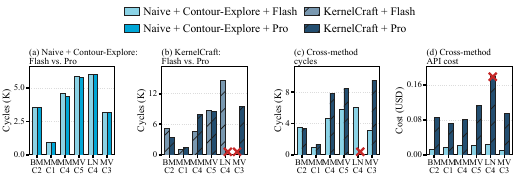}
  \par
  \setlength{\abovecaptionskip}{2pt}
\caption{Gemmini Flash/Pro ablation: (a) Naive + \sysexplore{}; (b) KernelCraft;
  (c, d) cycles and API cost for KernelCraft-Pro versus Naive + \sysexplore{} + Flash.
  Crosses denote failures. BM/MM/MV/LN:
  BatchMatMul/GEMM/GEMV/Linear.}
  \label{fig:model-ablation}
\end{figure}

\noindent\textbf{PLENA: generation success and kernel quality.}
Following the evaluation protocol used above, we compare KernelCraft and
Naive + \sysexplore{} with Sol on the 18 PLENA instances.
KernelCraft + Sol verifies 15 instances, versus 5 for DeepSeek-Flash
(median of three runs); Naive + \sysexplore{} + Sol verifies all 18.

Table~\ref{tab:plena-model-quality} compares performance on the 15 instances
solved by KernelCraft + Sol. Even after excluding its failures, KernelCraft's
geometric-mean speedup over \candidatezero{} is only $0.894\times$.
Naive + \sysexplore{} with the same model achieves
$1.262\times$ speedup on those same instances. Sol therefore
helps program-level generation solve previously unsuccessful workloads,
while the generated kernels collectively remain slower than the compiler
default and the mappings found through \sysexplore{}.

\begin{table}[t]
  \centering
\caption{PLENA coverage on 18 instances; geometric-mean speedup
  over \candidatezero{} on the 15 instances solved by KernelCraft + Sol. Higher is better.}
  \label{tab:plena-model-quality}
  \small
  \setlength{\tabcolsep}{4pt}
  \begin{tabular}{@{}lrr@{}}
    \toprule
    Method & Verified & Speedup $\uparrow$ \\
    \midrule
    \candidatezero{} & 18/18 & $1.000\times$ \\
    KernelCraft + Sol & 15/18 & $0.894\times$ \\
Naive + \sysexplore{} + Sol & 18/18 & $1.262\times$ \\
    \bottomrule
  \end{tabular}
\end{table}

\noindent\textbf{AccelOpt through \sysexplore{}.}
AccelOpt is an LLM optimization harness with a planner, executor, and
summarizer \cite{zhang2025accelopt}; we build on its public repository at
commit \texttt{47e29a7} \cite{acceloptcode2026}. Program-level AccelOpt edits
the kernel program, while AccelOpt through \sysexplore{} proposes mapping
intents through the interface of Section~\ref{sec:eval-gemmini}; both start
from \candidatezero{}.
On the 20 Gemmini instances with DeepSeek-Flash, AccelOpt through
\sysexplore{} returns a faster kernel than program-level AccelOpt on every
instance. Its kernels are $1.50\times$ faster in geometric mean, mean
disappointment falls from 36.8\% to 6.3\%, and it uses 588K instead of 620K
tokens per instance. Program-level AccelOpt returns a faster kernel than
the Gemmini library on 4 of the 20 instances, and AccelOpt through
\sysexplore{} on 17.
\sysexplore{} uses three runs and program-level AccelOpt uses one.

\noindent\textbf{PLENA: model choice within AccelOpt.}
We next compare DeepSeek-Flash and Sol within the AccelOpt harness on four
attention workloads: FlashAttention, MHA, GQA, and MQA, each at shape C3,
with one run per configuration.
AccelOpt + Sol tests whether a more capable model can complete the
program-level route without a compiler: it optimizes the kernels that
KernelCraft + Sol generates, whereas the other configurations start from
\candidatezero{}.
Within AccelOpt + \sysexplore{}, replacing DeepSeek-Flash with Sol improves
all four instances (Figure~\ref{fig:plena-model-accelopt}): mean disappointment falls from 1.48\% to 0.04\%,
mean tokens fall from 1.83M to 0.68M (62.7\% fewer), and geometric-mean
speedup over \candidatezero{} rises from $1.223\times$ to $1.241\times$.
AccelOpt + Sol raises the geometric-mean speedup of its KernelCraft-generated
kernels over \candidatezero{} from $0.771\times$ before optimization to
$0.934\times$ afterward, but retains 18.23\% mean disappointment and uses
8.01M tokens per instance. \textbf{With the same Sol model, AccelOpt + \sysexplore{}
achieves 32.8\% higher geometric-mean performance than this
program-level route at just 7.2\% of
AccelOpt's API cost (\$1.52 versus \$21.02 per instance).}
The benefit also holds for DeepSeek: \sysexplore{} reduces mean disappointment
from 17.92\% to 1.48\%, raises the geometric-mean speedup over
\candidatezero{} from $1.018\times$ to $1.223\times$ ($1.20\times$ faster
kernels), and uses 1.83M instead of 9.73M tokens per instance.

\begin{figure}[t]
  \centering
  \includegraphics[width=\columnwidth]{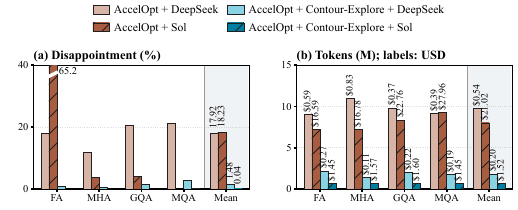}
  \par
  \setlength{\abovecaptionskip}{2pt}
\caption{AccelOpt on four PLENA attention instances (C3): (a)
  disappointment; (b) tokens per run (bars) and API cost in USD (labels).
  FA: FlashAttention.}
  \label{fig:plena-model-accelopt}
\end{figure}

Model capability thus remains useful on PLENA, both for generating correct
programs and for improving mapping search. The mapping interface preserves
that opportunity while providing a verified default that stronger models
can improve upon.

\endgroup

\subsection{Ablation of \sysext{}}
\label{sec:disc-ext-ablation}

We remove one mechanism of \sysext{} at a time and repeat the extension loop
of Section~\ref{sec:eval-plena} on four PLENA instances, FlashAttention C1,
SwiGLU C3, MHA C1, and FFN C1, with one run per configuration
(Table~\ref{tab:ext-ablation}). Without diagnosis, the extension agent
starts at the proposal phase and has no profiling tool. Without checks,
\sysext{} keeps every patch that applies and compiles \candidatezero{}, and
skips the correctness and effectiveness checks. Without failure feedback,
the history $H$ of rejected knobs is empty at every invocation.

\begin{table}[t]
  \centering
\caption{Ablation of \sysext{} on four PLENA instances: accepted knobs
  out of invocations and geometric-mean speedup over \candidatezero{}.
  $^{\dagger}$Excluding the incorrect knob.}
  \label{tab:ext-ablation}
  \begingroup\small
  \setlength{\tabcolsep}{5pt}
  \begin{tabular}{lrr}
    \toprule
    Configuration & Accepted & Speedup \\
    \midrule
    Full \sysext{} & 18/33 & $1.239\times$ \\
    \ablEvidenceTag No diagnosis & 9/25 & $1.143\times$ \\
    \ablEvidenceTag No failure feedback & 8/22 & $1.143\times$ \\
    \ablCheckTag No checks & 15/30 & $1.195\times^{\dagger}$ \\
    \ablTransferTag Leave-one-out & -- & $1.215\times$ \\
    Base surface & -- & $1.024\times$ \\
    \bottomrule
  \end{tabular}
  \endgroup
\end{table}

\noindent\ablEvidenceTag\textbf{Proposals need evidence.}
Diagnosis and failure feedback both tell the extension agent where to look:
the profile of the stalled program shows what limits it, and $H$ shows what
has already failed. Without either, fewer proposals pass the checks, the
loop ends earlier, and the final kernels are slower
(Table~\ref{tab:ext-ablation}); on FFN C1, no knob is accepted at all.
Without $H$, the agent also proposes rejected knobs again more often. To find useful
knobs, the agent therefore needs measured bottlenecks and past failures in
addition to the lowering code.

\noindent\ablCheckTag\textbf{Without checks, an incorrect knob reaches the final kernel.}
Without the checks, \sysext{} accepts a knob that skips writing an
intermediate result of FlashAttention C1 back to memory, although the
program later reads it. The program still passes the evaluator, because that
memory happens to hold valid values, and it is faster, so hill climbing
selects it for the final kernel. When memory is filled with NaN before
execution, its outputs contain NaN. Passing the evaluator during search does
not show that a knob is correct, so new knobs must be checked on additional
inputs before they enter the surface.

\noindent\ablTransferTag\textbf{Accepted knobs transfer across workloads.}
When the knobs that only the target instance contributed are removed from
the merged surface, the remaining knobs, all proposed for other instances,
still recover most of the gain of the instance's own extension loop. On the
two attention instances, they recover all of it. The knobs that \sysext{}
accepts are therefore reusable compiler capabilities that apply beyond the
workload they were proposed for, consistent with the merged surface in
Section~\ref{sec:eval-plena} matching or exceeding each instance's own
surface on 17 of the 18 instances.

 \section{Related Work}

\noindent KernelCraft generates ISA assembly for emerging accelerators, AutoComp
optimizes ISA-level C, and AccelOpt optimizes Trainium NKI kernels using
profiles and optimization memory
\cite{nie2026kernelcraft,hong2025autocomp,zhang2025accelopt}; GPU kernel
agents and benchmarks take the same program-level approach
\cite{sereda2025kforge,ouyang2025kernelbench,kernelevolve2026}. In these
workflows, the agent expresses its mapping choices through generated or
modified kernel code, which at the ISA level includes instruction calls,
address calculations, and data-movement schedules. In \sysexplore{}, the agent
sets knob values and the compiler generates the kernel; the search policies
of these harnesses can use the same interface, as our AccelOpt integration
shows (Section~\ref{sec:disc-accelopt}).

\noindent AutoTVM tunes the knobs of developer-written schedule templates
\cite{chen2018learning}, and Ansor and MetaSchedule generate schedule spaces
from fixed rules \cite{zheng2020ansor,shao2022metaschedule}. Timeloop
constructs per-layer mapspaces from hand-written architecture descriptions
and scores mappings analytically \cite{parashar2019timeloop}. In these
systems, the space is fixed before search, and a mapping outside it is
unreachable for any search policy. \sysexplore{} derives its mapping surface
from the RTL-lifted machine semantics and measures every mapping as a
generated kernel; when search stalls, \sysext{} adds knobs and their lowering
to the compiler, extending the surface itself (Section~\ref{sec:eval-vta}).

\noindent Concurrent to our work, CAKE~\cite{ye2026cake} proposes co-designing a GPU
compiler with kernel agents: agents write a typed, hardware-explicit
schedule IR, and the compiler returns localized diagnostics and evolves its
verifier rules and IR primitives when candidates expose missing
capabilities. Our work differs in three respects. First, CAKE targets NVIDIA
GPUs, for which large libraries of production kernels exist, and its
methodology builds on them: the IR is distilled from production kernels such
as those in FlashInfer and CUTLASS. Emerging accelerators have no such
corpus. Second, CAKE has no formal machine semantics; its primitives come
from production kernels and hardware documentation and are validated by
tests over its kernel corpus. \systemname{} instead derives its knobs from
the formal, RTL-lifted machine semantics of the target, and \sysext{} adds a
knob only after it passes legality checks against these semantics and the
numerical oracle. Third, CAKE's candidates are written by LLM agents,
whereas the mapping surface of \sysexplore{} can be searched by LLM agents and
by conventional policies such as hill climbing and GA alike.
 \section{Conclusion}

We present \systemname{}, a compiler-mediated framework for kernel
generation on emerging accelerators. \sysexplore{} connects search policies
to a reusable compiler through a workload-specific mapping surface: the
compiler completes partial mapping intents into legal plans, generates
target code, and returns verified correctness and performance feedback.
\sysext{} extends this surface when search stalls: an extension agent
implements a new knob and its lowering, and \sysext{} keeps the knob only if
it passes correctness and effectiveness checks.

The naive agent searching through \sysexplore{} returns a correct
kernel on every instance, whereas \kernelcraftname{} returns one on only
60\% of the 20 Gemmini instances and 5 of the 18 PLENA instances. On the
instances where \kernelcraftname{} succeeds, the \sysexplore{} agent's
kernels are $2.86\times$ (Gemmini) and $1.13\times$ (PLENA) faster than
\kernelcraftname{}'s, and \sysexplore{} spends 26.3\% and 83.4\% fewer
tokens per correct kernel. The \sysexplore{} agent's kernels are also
$1.17\times$ faster than Gemmini's library routine and $1.46\times$ faster
than those of the PLENA compiler. AccelOpt finds $1.50\times$ faster
kernels on Gemmini and
$1.20\times$ faster kernels on four PLENA attention instances when it
searches through \sysexplore{} instead of editing kernel programs. On 16 VTA instances,
\sysext{} extends the AutoTVM surface, and AutoTVM's XGBTuner finds
$4.14\times$ faster kernels on the extended surface; on PLENA, the extended
surface makes the kernels found by hill climbing $1.20\times$ faster.
These results show that partial mapping specification
with compiler completion lets neural and conventional search share one
verified interface, and that this interface can be extended when search
stalls.

A future direction is recursive improvement across data, models, and
harnesses, inspired by MetaRSI~\cite{tan2026metarsi}. The verified kernels,
mapping decisions, and measured performance produced by \sysexplore{},
together with the knobs accepted by \sysext{}, could serve as post-training
data. A post-trained model could then search more efficiently and propose
further knobs; once these knobs pass the checks of \sysext{}, they would
expand subsequent searches and produce further training data.
 
\bibliographystyle{ACM-Reference-Format}
\bibliography{refs}

\end{document}